# Applicability of Electrical Conductivity Ratio Method to Complicated Band Structure and the Carrier Scattering Mechanisms of SnSe

Pan Ren1, Junling Gao1, Guiying Xu1*, Bohang Nan1, Tao Guo1, Quanxin Yang1, Fanchen Meng2, Myles McKenna3, Jian He3* and Sitong Niu1

1. Beijing Municipal Key Lab of Advanced Energy Materials and Technology, School of Materials Science and Engineering, University of Science and Technology Beijing, Beijing 100083, China.
2. National Laboratory for Infrared Physics, Shanghai Institute of Technical Physics, Chinese Academy of Science, Shanghai 200083, China.
3. Department of Physics and Astronomy, Clemson University, Clemson, SC 29634-0978, USA.
*Corresponding author: xugy@mater.ustb.edu.cn

**Abstract:** The electrical conductivity ratio (ECR) method can be used to analyze carrier scattering mechanism without the need of magenetic transport measurements. In this work, the applicability of the ECR method in the analysis of complex energy band structures is discussed. Combined with the thermoelectric properties of SnSe, the feasibility using ECR method of ideal single band transport model to study the carrier scattering mechanism of semiconductor materials with complicated band structure is studied. The results indicate that ECR method is not only applicable to idea band structure as reported before but also to the complicated band structure, such as anisotropic band structure with one or multi-valley parabolic band, and isotropic band structure with one or multi-valley nonparabolic band. The analysis results of the carrier scattering mechanism of single crystal SnSe by ECR method using ideal single-band model agree with the carrier mobility-temperature dependence in the non-phase transition temperature range. Therefore, it is feasible to study the carrier scattering mechanism of semiconductor materials with complicated band structure by ECR method using ideal single band transport model. The carrier scattering mechanisms along three direction of single crystal SnSe are different because of its anisotropic crystal structure. Dislocation scattering (DS) mechanism has very similar manifestation to charged impurity scattering (CIS). The difference between DS and CIS is that DS is always accompanied by polar optical phonon scattering (POP), such as the carrier scattering mechanism (CSM) of SnSe along b axis direction at 573 – 823K. The difference between ionized impurity scattering and dislocation scattering can be more easily distinguished by the ECR method than the carrier mobility-temperature dependent method. The reason is that in the process of ECR analysis, it is always necessary to consider the role of multiple scattering mechanisms. DS and POP might be one of the approaches to improve thermoelectric property because the scattering factor for DS (3/2) or POP (1/2) is larger than that of acoustic phonon scattering (APS) and alloy scattering (AS). For polycrystalline SnSe, the carrier scattering mechanisms varies with the crystal structure and the temperature. For instance, in the case of vacuum-synthesized polycrystalline SnSe, both before and after the phase transition, acoustic phonon scattering is the dominant scattering mechanism. However, during the phase transition process (473K < T < 673K), polarization scattering is the dominant scattering mechanism. This result also indicates that the end temperature of the phase transition for this polycrystalline SnSe is lower than that of single-crystalline SnSe (823K). The nano-crystalline SnSe prepared by high-pressure sintering mainly exhibits ionized impurity scattering in the temperature range of 309K ≤ T ≤ 425K. When 425K < T ≤ 525K, dislocation scattering becomes the dominant mechanism. Moreover, its phase transition is completed at 525K. This indicates that the ECR method can better reflect the variation of carrier scattering mechanisms with temperature compared to the carrier mobility temperature dependence method.

## 1. Introduction

Crucial in many regimes of science, engineering and technology, the study of carrier scattering mechanism (CSM) provides invaluable insights into the interplay between charge carriers and other quasi-particles (phonons, defects etc.) at the microscopic scale [1-7]. To date, the mainstream or traditional analysis method of CSM still rely on the temperature (T) dependence of carrier mobility μ(T). It can be simply referred to as the "Carrier Mobility Temperature Dependence" (CMTD) method. It is calculated from the experimentally measured Hall coefficient $R_H(T)$ and electrical conductivity σ(T) using the relation μ(T) = σ(T)$R_H$(T) [3-7]. In the context of relaxation time approximation, the relaxation time $\tau \propto \tau_0 E^\gamma$, where $\tau_0$ is a sample dependent constant, E is the energy of charge carrier, and $\gamma$ is the carrier scattering factor [8]. For charged (ionized) impurity scattering (CIS) and dislocation scattering (DS), r = 3/2; for polar optical phonon scattering (POP) [8], r = 1/2; for acoustic phonon scattering (APS), alloy scattering (AS) [8], and grain boundary barrier scattering (GBS) [9-11], r = -1/2. Importantly, the μ(T) exhibits a power law dependence on T : $\mu(T) \propto \mu_0 T^{\gamma'}$, where $\mu_0$ is a sample dependent constant, $T^{3/2}$ for charged impurity scattering and dislocation scattering, $T^{-3/2}$ for acoustic phonon scattering and polar optical phonon scattering, $T^{-1/2}$ for alloy scattering, and T0 for charge neutral impurity scattering [8]. These characteristic temperature dependences are used as the indicator to pinpoint the underlying dominant CSM, although these dependences are all subject to simplification conditions, which are rarely perfectly satisfied in practice. However, this analysis procedure is severely limited by the measurement of Hall coefficients under high temperatures, the measurement of materials containing magnetic elements, as well as the strict requirements for the high uniformity of the composition and structure of the samples. It is thus highly desired to have an alternative analysis method of CSM without the need of galvanomagnetic transport measurements.

In light of the fact that both electrical conductivity and Seebeck coefficient depend on the electron band structure and carrier scattering mechanism, based on the temperature dependence of the ratio of electrical conductivity (ECR) for materials with single-valley parabolic energy band structure, an analysis method for CSM was designed [1]. It is abbreviated as the ECR method. This method pinpoint the dominant scattering mechanism by comparing the electrical conductivity ratios σ(T)/σ($T_0$) vs. temperature T in the theoretical calculation and experimental results based on thermoelectric (TE) property measurement. The closest one to the experimental value is the dominant carrier scattering mechanism. In this ECR procedure, the demanding Hall coefficient measurement is no longer needed. Compared to the CMTD method, the ECR method is more tolerant of sparse data points, more stable at high temperatures, and suitable for rare earth element containing materials. Another advantage of the ECR method over the CMTD method is that the measurement accuracy and reproducibility of Seebeck coefficient is higher compared to the Hall coefficient test. Since Seebeck coefficient measurement has lower requirements for the uniformity of the sample's composition and structure. Therefore, using Seebeck coefficient to analyze the scattering mechanism of samples is at least more accurate than using Hall coefficient. However, it is essential to understand the applicability of ECR method to the materials with complicated band structure.

Although many materials have complex band structures, their properties can be explained by theoretical models with simple band structures. Therefore, the applicability of the ECR method in materials with complex band structures should be first examined. Then, by combining the thermoelectric properties of SnSe, it is discussed whether the ECR method of the single-valley parabolic band model could be used to describe the carrier scattering mechanism of materials with complex band structures. There are three reasons to choose the carrier scattering mechanism of SnSe as the research object. Firstly, SnSe is known to have anisotropic, non-parabolic and multivalley electron band structure [13]. Secondly, SnSe formed by environmentally friendly and naturally abundant elements, has attracted intense attention for its high dimensionless thermoelectric (TE) figure of merit ZT [12-29], ZT=$\alpha^2\sigma T/\kappa$, where α, σ, κ and T are the Seebeck coefficient, electrical conductivity, thermal conductivity and absolute temperature respectively. There are two major motivations for this work. At first, the ECR method is valid when the effective mass is temperature independent. Besides, SnSe is a fascinating material holding promise in applications in photovoltaics [30], and lithium ion batteries [31-32]. Thirdly, up to now, the reported CSM in SnSe derived from traditional galvanomagnetic transport measurements varies greatly [12-18]. In single crystalline SnSe, the dominant scattering mechanism was reported to be charged impurity scattering [12] or acoustic phonon scattering in the entire temperature range studied [13], or evolving from charged impurity scattering between 141k and 208k, to acoustic phonon scattering between 208 K and 468K, and to a complex carrier scattering mechanism between 468 K and 553K [14]. In particular, the μ(T) increases with temperature around 600-800K, where the electron-phonon coupling dominates and decreasing μ(T) is expected [13, 33]. The situation is more complicated in polycrystalline SnSe samples, the dominant CSM was found to be intimately related to the defects, synthesis recipe and temperature. For example, the dominant scattering mechanism at room temperature varies from charged impurity scattering in pristine SnSe samples fabricated by high pressure sintering (HPS) method [16-17], to acoustic phonon scattering in Sn0.99Zn0.01Se fabricated by hot pressing (HP) method [15], and to grain-boundary potential barrier scattering in Sn0.99K0.01Se fabricated by mechanical alloying (MA) followed by spark plasma sintering (SPS) [18]. Meanwhile, the dominant CSM evolves from charged impurity scattering or grain-boundary potential barrier scattering at low temperatures to acoustic phonon scattering at high temperatures [15-18]. All of these results were obtained from the traditional analysis method based on galvanomagnetic transport measurements. Some immediate questions arise as to whether these discrepancies are due to different synthesis recipes, defects, or the uncertainty of Hall coefficient measurements under higher temperatures. As a cross-validation, the ECR method and the published data of the aforementioned typical samples (including single-crystal SnSe [13] and polycrystalline SnSe [16,19] within the temperature range of 300 K to 873 K) are utilized to analyze the carrier scattering mechanisms of the corresponding samples. The reason for citing reference [13] is that it was published in the Nature journal, making it more representative and having sufficient influence. The reasons for citing references [16 and 19] are that both of these two papers are the research results of our research group, meaning the data is highly reliable. That is to say, the thermoelectric performance data of SnSe reported in these three references meet the requirements of representativeness and reliability. Therefore, the results of using the ECR method to analyze the carrier scattering mechanism of single-crystal and polycrystalline SnSe based on the thermoelectric performance data cited above have high reliability and representativeness.

Based on the above discussion, the first step here is to investigate the applicability of the ECR method to materials with complex energy band structures. Next, the feasibility using the ECR method in conjunction with the spherical energy surface single-band thermoelectric theory model to study the carrier scattering mechanism of SnSe with complex energy band structure is discussed. Finally, the ECR method is used to study the variation law of the carrier scattering mechanism of single-crystal and polycrystalline SnSe with temperature.

## 2. The applicability of ECR procedure to complex band structure

Before using the ECR method to analyze the transport data of SnSe it is imperative to address the applicability of ECR procedure to complex electron band structure (e.g., anisotropy, multivalley, and non-parabolicity).

### 2.1. Anisotropic band structure with single-valley parabolic band

For anisotropic band structure (ellipsoidal constant energy surface) with parabolic band, assuming that the effective mass along the three crystallographic axes are $m_1$, $m_2$ and $m_3$, and the relaxation time τ is isotropic, the mean inertial effective mass or electrical conductivity effective mass mc* along any direction can be described by formula (1) [8]:

$$\frac{1}{m_c^*} = \frac{1}{3}(\frac{1}{m_1} + \frac{1}{m_2} + \frac{1}{m_3}) = \frac{m_1 m_2 + m_1 m_3 + m_2 m_3}{3 m_1 m_2 m_3} \qquad (1)$$

The carrier mobility μ can be described as [8]:

$$\mu = \frac{2e}{3}(r+\frac{3}{2})(k_B T)^r \frac{\tau_0}{m_c^*} \frac{F_{r+1/2}(\eta)}{F_{1/2}(\eta)} \tag{2}$$

$\tau_0$ can be described as [1]:

$$\tau = \tau_0 E^r \tag{3}$$

$$\tau_0 = \tau_0'(m^*)^{c_1} T^{c_2} \tag{4}$$

By substituting formulas (1) and (4) into formula (2), the mobility expression can be written as:

$$\mu = \frac{2e}{3}(r+\frac{3}{2}) k_B^r T^{c_2+r} \left[\frac{3m_1 m_2 m_3}{(m_1 m_2 + m_1 m_3 + m_2 m_3)}\right]^{c_1-1} \frac{F_{r+1/2}(\eta)}{F_{1/2}(\eta)} \tag{5}$$

Where τ, e, r, k$_B$, and η are the the relaxation time, electronic charge, scattering factor, Boltzmann constant, and the reduced Fermi level measured from the band edge, respectively. c$_1$ and c$_2$ are constants independent on temperature. Fj(η) is the Fermi-Dirac integral of order j.

In this case, the density-of-state effective mass m$_N$* is different from m$_c$*. m$_N$* can be described as equation (6):

$$m_N^* = (m_1 m_2 m_3)^{1/3} \tag{6}$$

For the semiconductor with single-valley parabolic band, the carrier concentration n and the electrical conductivity σ can be described in formula (7) [8] and (8), respectively.

$$n = \frac{4\pi(2m_N^* k_B T)^{3/2}}{h^{3/2}} F_{1/2}(\eta) = \frac{4\pi(2k_B T)^{3/2}(m_1 m_2 m_3)^{1/2}}{h^{3/2}} F_{1/2}(\eta) \tag{7}$$

$$\sigma(T) = ne\mu = \frac{16\sqrt{2}\pi e^2 \tau_0' 3^{c_1}(m_1 m_2 m_3)^{c_1-1/2}(k_B T)^{r+3/2} T^{c_2}}{h^{3/2}(m_1 m_2 + m_1 m_3 + m_2 m_3)^{c_1-1}}(r+\frac{3}{2}) F_{r+1/2}(\eta) \tag{8}$$

If assuming the m$_i$ (i = 1, 2, 3) is constant in certain temperature range, the ratio of electrical conductivity at temperature T (σ(T)) to the electrical conductivity at certain temperature T$_0$ (σ(T$_0$)), or σ(T)/ σ(T$_0$), will only depend on the carrier scattering factor r, as shown in formula (9).

$$\sigma(T)/\sigma(T_0) = \frac{(T)^{r+3/2+c_2} F_{r+1/2}(\eta_T)}{(T_0)^{r+3/2} F_{r+1/2}(\eta_{T_0})} \tag{9}$$

In other words, the ECR method is suitable for the anisotropic band structure (ellipsoidal constant energy surface) with single-valley parabolic band.

**2.2. Anisotropic band structure with multi-valley parabolic band**

Assuming that the number of the multi-valley parabolic band is N, based on formula (8), the electrical conductivity can be expressed by the formula (10):

$$\sigma(T) = \sum_i^N \sigma_i = \frac{16\sqrt{2}\pi e^2 \tau_0' 3^{c_1}(k_B T)^{r+3/2} T^{c_2}}{h^{3/2}}(r+\frac{3}{2})\sum_i^N \frac{(m_{1i} m_{2i} m_{3i})^{c_1-1/2}}{(m_{1i} m_{2i} + m_{1i} m_{3i} + m_{2i} m_{3i})^{c_1-1}} F_{r+1/2}(\eta_i) \tag{10}$$

where σ$_i$ presents the contribution of valley i to the electrical conductivity, η$_i$ is the reduced Fermi level measured from the band edge. If assuming the m$_i$ is constant, the C$_i$ shown in formula (11) is constant, the σ(T)/σ(T$_0$) will only depend on carrier scattering factor (r), as shown in formula (12), and ECR method is applicable.

$$C_i = \frac{(m_{1i}m_{2i}m_{3i})^{c_1-1/2}}{(m_{1i}m_{2i}+m_{1i}m_{3i}+m_{2i}m_{3i})^{c_1-1}} \tag{11}$$

$$\sigma(T)/\sigma(T_0) = \frac{T^{r+3/2+c_2}\sum_i^N F_{r+1/2}(\eta_{iT})}{T_0^{r+3/2+c_2}\sum_i^N F_{r+1/2}(\eta_{iT_0})} \tag{12}$$

**2.3. Isotropic band structure with single-valley non-parabolic band.**

For isotropic band structure (spherical constant energy surface) with single-valley non-parabolic band, considering both the relaxation time approximation and the mean free path approximation, according to the reports, the electrical conductivity can be described by formula (13) [34,35]:

$$\sigma(T) = \frac{2ge^2\tau_0 2\sqrt{2m_0}}{3m^*h^{3/2}} \int_0^\infty \left(\frac{E}{E_0}\right)^r \frac{(1+aE)^{3/2}}{1+2aE}\left(\frac{\partial f}{\partial E}\right)dE \tag{13}$$

Where g, $\tau_0$, h, $m^*$, $m_0$, a, E, $E_0$, r and f are the constant of the valley degeneracy, the constant in the expression of relaxation time approximation, the reduced Planck constant, the electron dimensionless effective mass, the electron mass, the non-parabolicity parameter, the energy measured with respect to the conduction band edge, the constant E0 = $k_B$300K, the parameter related to the carrier scattering mechanism, and the Fermi function.

The expression for the electrical conductivity, given in (13), contains six implicit assumptions: (i) that quantization effects can be neglected, (ii) that the conductivity can be treated as being isotropic; (iii) that deviations of the true distribution function from the equilibrium Fermi distribution function $f_0$ are small enough to justify the use of the Boltzmann transport equation in its linearized form; (iv) that the material is sufficiently n-doped to neglect both valence band and other conduction band contributions; (v) the relaxation time is approximately valid.; and (vi) that the relaxation time τ can be approximated as τ=$\tau_0$(E/$E_0$)β with a relaxation time constant $\tau_0$ and a normalization constant $E_0$= $k_B$300K necessary for dimensionality reasons. The coefficient β describes the energy dependence of the relaxation time which, via the temperature dependence of the Fermi function $f_0$, implicitly causes a temperature dependence of τ of approximately (T/300K)β. This energy dependence is often neglected and instead relaxation time is made explicitly temperature-dependent.

If assuming the $\tau_0$, $m^*$ is constant, the σ(T)/σ($T_0$) will only depend on carrier scattering factor (r), as shown in formula (14), and ECR method is applicable.

$$\sigma(T)/\sigma(T_0) = \frac{\left[\int_0^\infty \left(\frac{E}{E_0}\right)^r \frac{(1+aE)^{3/2}}{1+2aE}\left(\frac{\partial f}{\partial E}\right)dE\right]_{T=T}}{\left[\int_0^\infty \left(\frac{E}{E_0}\right)^r \frac{(1+aE)^{3/2}}{1+2aE}\left(\frac{\partial f}{\partial E}\right)dE\right]_{T=T_0}} \tag{14}$$

Therefore, for isotropic band structure (spherical constant energy surface) with single-valley non-parabolic band, when the effective mass is constant in some temperature range, the σ(T)/σ($T_0$) only depends on carrier scattering factor, as shown in formula (14). Similarly to the proving process of the applicability of ECR to anisotropic band structure with multi-valley parabolic band, ECR is applicable to isotropic band structure with multi-valley nonparabolic band.

From the above discussion, it can be concluded that the ECR method is applicable to all the complex band structure materials discussed previously.

**3. Procedure of ECR Analysis**

According to the literature [36], SnSe has three valence bands. The energy and shape of the two valence bands with high energy are almost identical. The energy of the third valence band is much lower than that of the two above, and its influence on the performance of SnSe can be negligible. Considering the characteristics of the ECR method (see equation (7)), the effect of multiple valence bands on the performance of SnSe can be approximately ignored. In other words, the carrier scattering mechanism can be analyzed by ECR method using the single band model. In addition, although SnSe has an anisotropic crystal structure, the variation of thermoelectric properties of semiconductor materials can generally be explained with an isotropic thermoelectric model. Therefore, the scattering mechanism of SnSe is studied by the ECR method using single-valley parabolic band model with isotropic band structure (spherical constant energy surface).

Table 1. Transport behaviors for typical carrier scattering mechanisms in a non-degenerate parabolic band structure [a].

| η | Carrier scattering mechanism | τ (Carrier relaxation time) [b] | α (Seebeck coefficient) | σ (Electrical conductivity) [c] |
|---|---|---|---|---|
| Nonde-generate limit (η< 0) | Acoustical phonon scattering (APS) | $\tau_0 T^{-1} E^{-1/2}$ | $\frac{k_B}{e}(2-\eta)$ | $\sigma_0 \exp(\eta)$ |
| | Alloy scattering (AS) | $\tau_0 E^{-1/2}$ | $\frac{k_B}{e}(2-\eta)$ | $\sigma_0 T \exp(\eta)$ |
| | Polar optical phonon scattering (POP) | $\tau_0 E^{1/2}$ | $\frac{k_B}{e}(3-\eta)$ | $\sigma_0 T^2 \exp(\eta)$ |
| | Charged impurity scattering (low temperature) (CIS) | $\tau_0 h(T) E^{3/2}$ | $\frac{k_B}{e}(4-\eta)$ | $\sigma_0 h(T) T^3 \exp(\eta)$ |
| | Dislocation scattering (DS) | $\tau_0 E^{3/2}$ [27,28] | $\frac{k_B}{e}(4-\eta)$ | $\sigma_0 h(T) T^3 \exp(\eta)$ |

a. Here $\eta = E_F/k_B T$; the zero point of $E_F$ is set at the band edge.
b. For acoustical phonon and alloy scattering $\tau_0 \propto (m^*)^{-3/2}$, polar optical phonon scattering $\tau_0 \propto (m^*)^{-1/2}$, charged impurity scattering $\tau_0 \propto (m^*)^{1/2}$ [8], grain boundary potential barrier scattering $\tau_0 \propto (m^*)^{1/2}$ [9-11], dislocation scattering $\tau_0 \propto (m^*)^{1/2}$ as shown in formula above (4) [37,38].
c. For all six mechanisms $\sigma_0 \propto \tau_0 (m^*)^{1/2}$ [8].

In the ECR analysis procedure [1], it is assumed a temperature independent band effective mass $m_b^*$ in the case of a single-valley parabolic band model. Specific scattering mechanisms considered herein are acoustical phonon scattering, alloy scattering, polar optical phonon scattering, charged impurity scattering and dislocation scattering. It is noted that the analysis results of carrier scattering mechanism using degenerate semiconductor series formula or non-degenerate semiconductor series formula in electrical conductivity ratio (ECR) method are consistent. Hence, for convenience, the nondegenerate model (formulas) is adopted to calculate the reduced Fermi energy level (η) and electrical conductivity ratio $\sigma(T)/\sigma(T_0)$, where $T_0$ is a reference temperature. Generally, the $T_0$ is chosen at which the Seebeck or electrical conductivity changes sharply [1]. Multiple $T_0$ can be set to ensure within each temperature situation the $m_b^*$ can be regarded as a constant.

The ECR analysis procedure is implemented as following. First, the temperature-dependent Fermi energy η is derived from the experimental values of the Seebeck coefficient α vs. T in the case of the model selected (cf. Table 1). Secondly, the derived η value is used to calculate $\sigma(T)/\sigma(T_0)$ (ECR) in the case of the model selected (cf. Table 1). Thirdly, plot the temperature function curves of $\sigma(T)/\sigma(T_0)$ for various scattering mechanisms, compare them with the experimental results, and finally pinpoint the dominant CSM.

As shown in table 1, or according to the literature, the carrier scattering factor of dislocation scattering is 3/2 and carrier mobility temperature dependence ($\mu \propto T^{3/2}$) similar to those of charged impurity scattering (low temperature upon suitable simplification [37,38]. One way is needed to distinguish between them. It can be described below.

To begin with, the carrier momentum relaxation time ($\tau_m$) for dislocation scattering can be described by the formula (15)[37,38]:

$$\tau_m = \frac{8\varepsilon^2\varepsilon_0^2 d^2 m^2}{Ne^4\varphi^2 L_D}(v_e^2 + h^2/4m^2\tau_d^2)^{3/2} \tag{15}$$

where $\varepsilon$ is the dielectric constant, $\varepsilon_0$ is the dielectric constant in vacuum, d is the distance between two imperfections on dislocation line, φ is the probability of defects on a dislocation line, N is the amount of dislocation lines per unit area, $L_D$ is Debye length, m is carrier effective mass, $v_e$ is carrier speed, h is Planck constant, and $\tau_d$ is dielectric relaxation time. If E is the energy of the carrier, formula (16) can be expressed as formula (16):

$$\tau_m = \frac{8\varepsilon^2\varepsilon_0^2 a^2 m^{1/2}}{Ne^4\varphi^2 L_D}(mv_e^2 + h^2/4m\tau_d^2)^{3/2} = \frac{8\varepsilon^2\varepsilon_0^2 a^2 m^{1/2}}{Ne^4\varphi^2 L_D}E^{3/2} \tag{16}$$

$$E = mv_e^2 + h^2/4m\tau_d^2 \tag{17}$$

In the maximum temperature limit, the second item in the parentheses of formula (16) can be neglected, the carrier mobility can be described as [37,38]:

$$\mu = \frac{30\sqrt{2\pi}\varepsilon^2\varepsilon_0^2 a^2}{e^3\varphi^2 L_D m^{1/2}}\frac{(k_B T)^{3/2}}{N} \tag{18}$$

Where $k_B$ and T are Boltzmann constant and temperature respectively.

Besides, the second item in parentheses of formula (15)~(17) is dielectric relaxation time ($\tau_d$), which represents the effect of polar optical phonon scattering on the CSM, and dislocation scattering is always accompanied by polar optical phonon scattering. This result indicates that dislocation scattering not only introduces dislocation scattering resulting from lattice defects, but also incorporates polarization scattering caused by the uneven distribution of charges. However, the above characteristics of dislocation scattering cannot be given clearly when judging the carrier scattering mechanism by the carrier mobility temperature dependent relation or CMTD method according to formula (18). Therefore, charged impurity scattering and dislocation scattering cannot be separated by CMTD method [39,40].

## 4. Results and discussion
### 4.1. Single crystalline SnSe

Fig. 1 shows the Seebeck coefficient (α) and electrical conductivity (σ) data measured along three crystallographic directions (a, b, and c axis) of single crystalline SnSe samples reported by Zhao [13], polycrystalline SnSe fabricated by vacuum melting (SnSe-VM) [19], and polycrystalline SnSe fabricated by HPS (SnSe-HPS) [16]. It is found that single crystalline SnSe has higher Seebeck coefficient and electrical conductivity than polycrystalline SnSe almost over the entire temperature range measured except the Seebeck coefficient of SnSe-HPS at 625K – 823K.

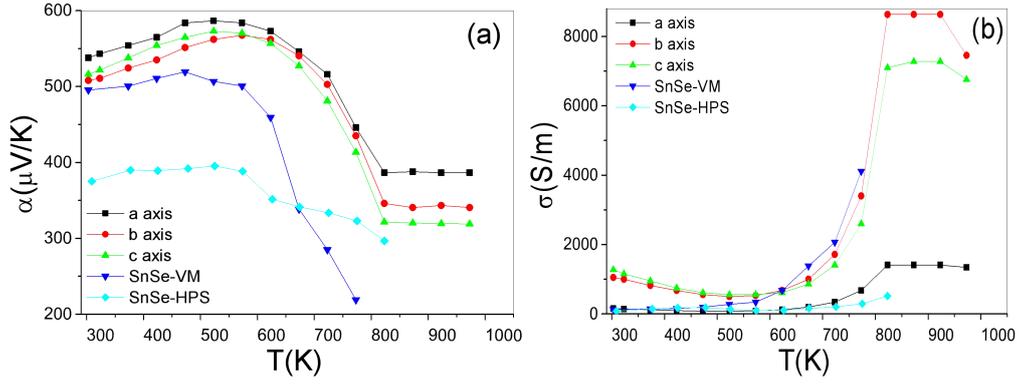

Figure 1. (a)Seebeck coefficients and (b) electrical conductivities of three directions (a, b, and c axis ) of single crystal reported by Zhao[13], SnSe fabricated by vacuum melting (SnSe-VM), and SnSe fabricated by by HPS (SnSe-HPS).

As shown in Fig. 1, the Seebeck coefficients and electrical conductivities along a, b, and c axes of SnSe single crystal reported by Zhao et al. [13] have similar temperature dependences. Because the Seebeck coefficient and the electrical conductivity exhibits maximum and minimum near 523K (along a and c axis) or 573K (along b axis), and almost remain constant above 823K, respectively; 300 K, 523 K (along a and c axis), or 573K (along b axis), and 823K along three directions were adopted as $T_0$. Besides, grain-boundary potential barrier scattering mechanism was not considered because there is no grain boundary in single crystal SnSe.

The results of ECR analysis along a, b and c axis are shown in Fig. 2 (a), (b) and (c) respectively. Notably, the derived CSM along three major crystallographic directions are different. Along a axis (cf. Fig. 2 (a)), judging from the calculated line closest to the experimental result (EXP), it indicates that the dominant CSM is acoustical phonon scattering (APS) from 300K to 523K, polar optical phonon scattering (POP) when 523K<T< 823K, and acoustic phonon scattering (APS) after 823K respectively. As for the b axis, the main CSM is same as along a axis except for the temperature range of 573K to 673K, which is dominated by charged impurity scattering (CIS). The dominant CSM along c axis direction is different from the other 2 directions: it is independent on temperature and acoustic phonon scattering (APS) as the dominant CSM over the whole temperature range. The analysis results of the ideal single-band model in the non-phase transition temperature range obtained by the ECR method are consistent with the mobility temperature dependence [41]. Therefore, it is feasible to study the carrier scattering mechanism of semiconductor materials with complicated band structure by ECR method using ideal single band transport model. The very interesting observation is that charged impurity scattering (CIS) plays an important role in the carrier scattering process for single crystalline SnSe along b axis at 573K<T<673K. This is an uncommon phenomenon because it is usually thought that the charged impurity scattering (CIS) mechanism plays a dominant role in polycrystalline materials at low temperatures. The reasons or the mechanism can be discussed below.

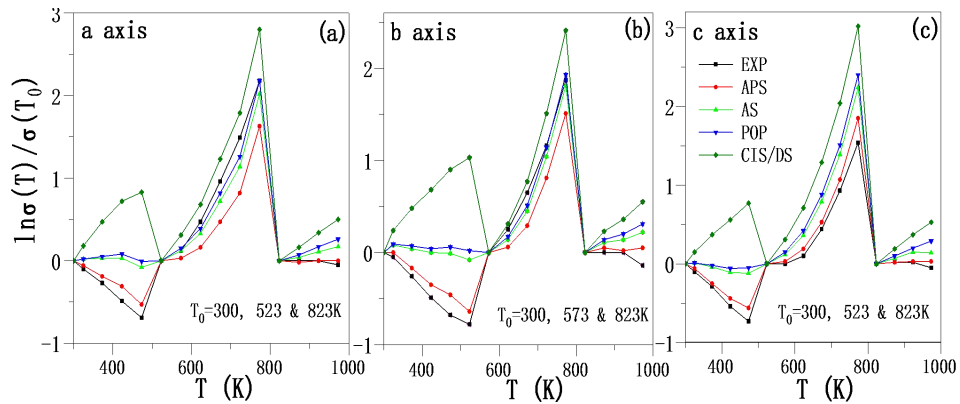

Figure 2. Calculation results for SnSe single crystal along a (a), b (b), and c (c) axis direction, where line EXP corresponds to experimental ratio of electrical conductivity σ(T) to the electrical conductivity at $T_0$ K vs. temperature T, and lines acoustical phonon scattering (APS), alloy scattering (AS), polar optical phonon scattering (POP), charged impurity scattering (CIS) and dislocation scattering (DS) are the predicted ratios vs. T based on APS, AS, POP, CIS and DS respectively.

At first, SnSe is binary IV–VI group semiconductor with anisotropic crystal structure (cf. Fig. 3), and shows a layered structure crystallized in the orthorhombic Pnma space group (GeS type) from room temperature to 807K that is derived from the NaCl-type structure. The tin atoms and selenium atoms form a zig-zag structure, which is a distorted octahedron [42-45]. Therefore, the structural anisotropy should have strong effect on the transport property [13].

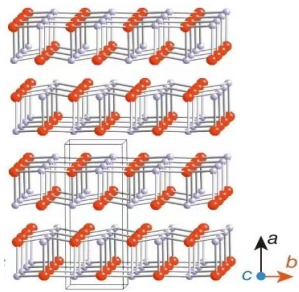

Figure 3. SnSe crystal structure with a space group of Pnma. The primitive cell is shown as a black line lattice. Red larger balls are Se atoms and gray smaller balls are Sn atoms. The directions of three axes (a, b and c) are also shown as black, red and blue arrows, respectively [13].

Secondly, pristine SnSe is a native p-type semiconductor, so the majority carriers are holes which move generally along and near $Se^{2-}$ surface. When the holes move along a or b axis direction, or TE properties measured along a or b axis direction, they will meet $Sn^{2+}$ and $Se^{2-}$ or be scattered by ionized $Sn^{2+}$ and $Se^{2-}$ at the same time. Therefore, the acoustic phonon scattering is normally the major CSM at the range of 300K to 523K. Above 523K, there will be a phase transition from α SnSe to β SnSe along a axis direction [33, 46], the lattice parameters a, b, and c are changed from 0.4436, 0.4162 and 1.149 nm for α SnSe to 0.4310, 1.1705 and 0.4318 nm for β SnSe respectively. During the pre-stage of the phase transition, the center of positive charges are separated from the center of negative charges. That is the polarization is formed between $Sn^{2+}$ and $Se^{2-}$ because of atoms or ions movement induced by phase transition. As the result, the polar optical phonon scattering (POP) becomes the dominant CSM during phase transition. If dislocation scattering (DS) induced by phase transition is stronger than polar optical phonon scattering (POP), such as b axis direction at the temperature range of 573K to 673K, charged impurity scattering (CIS) or rather dislocation scattering (DS) will become dominant. The evidences for the polarization are the rapidly changing electrical conductivities and Seebeck coefficients at 523K to 823K or 573K to 823K along a or b axis respectively. After phase transition, the polarization disappears and the acoustical phonon scattering (APS) becomes the dominant CSM again. Thirdly, it can be demonstrated that dislocation scattering (DS) is dominant along b axis at 573K-673K. Because polar optical phonon scattering (POP) is bound to occur when dislocation scattering (DS) is the dominant scattering mechanism, as shown in equations (15) and (16). According to Figure 2 (b), at 573K<T<673K, except dislocation scattering, polar optical phonon scattering is also very close to the experimental value. While dislocation scattering is the dominant scattering mechanism, polar optical phonon scattering also plays an important role. It meets the requirement that dislocation scattering is the dominant scattering mechanism. Therefore, it is phase transition that induces dislocation scattering instead of charged impurity scattering becoming the dominant CSM, although it shows up as charged impurity scattering. This result is consistent with experimental result of report [33]. As for c axis, the surface which holes move along only consists of $Se^{2-}$, in other words, the holes cannot be scattered by $Sn^{2+}$. As the result, acoustical phonon scattering (APS) is dominant in whole temperature range. Another result obtained from the analysis above, the phase transition of α SnSe to β SnSe begin at 523K (along a and c axis) or 573K (along b axis).

**4.2. Polycrystaline SnSe**

There are many reports on the electrical conductivity and Seebeck coefficient of polycrystalline pristine SnSe samples. The data of polycrystalline SnSe samples synthesized by vacuum melting (VM) method [19] and those prepared by high-pressure sintering (HPS) method [16] are compared. Two samples are hereinafter referred to as SnSe-VM and SnSe-HPS respectively. As shown in Fig. 1, the Seebeck coefficient of SnSe-VM exhibits maxima at 473K. Therefore, 303K and 473K were taken as T0 values. The analysis results are presented in Fig. 4(a). Except for the regime: 473K<T<673K, where polar optical phonon scattering (POP) is closest to line EXP, or polar optical phonon scattering (POP) is the dominant scattering mechanism, acoustical phonon scattering (APS) seems to dominate in other temperature range although the description is less satisfactory. This result is similar to that of single crystal SnSe along a axis direction. The reason is SnSe-VM has extremely high [100] orientation [19].

One special phenomenon is that the temperature for phase transition temperature (473K <T< 673K) is lower than that (523K < T < 823K) for single crystal SnSe along a axis direction. The reason is that polycrystal has smaller particle size, larger grain boundary area and much more crystal defects than single crystal SnSe. All of these crystal defects in polycrystal decrease the temperature for phase transition.

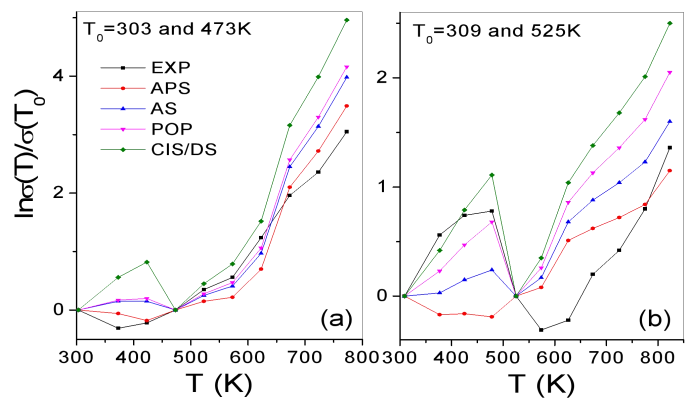

Figure 4. Calculation results for polycrystal SnSe fabricated by(a) VM (SnSe-VM) and (b) HP (SnSe-HPS) respectively, where line EXP corresponds to experimental ratio of electrical conductivity σ(T) to the electrical conductivity at T0 K vs. temperature T , and lines acoustical phonon scattering (APS), alloy scattering (AS), polar optical phonon scattering (POP), charged impurity scattering (CIS) and dislocation scattering (DS) are the predicted ratios vs. T based on APS, AS, POP, CIS and DS respectively.

As shown in Fig.1(a), the Seebeck coefficient of SnSe-HPS shows a maxima at 525K. Therefore, 309K and 525K are taken as T0. The analysis results are shown in Fig. 4 (b). The carrier scattering mechanism varies with temperature too. When 309K≤T≤425K, line charged impurity scattering (CIS)/ dislocation scattering (DS) is closest to line EXP, suggesting charged impurity scattering (CIS)/dislocation scattering (DS) is dominant in this situation. The result is consistent with the early report [16-17]. When 425K< T≤525K, line POP is close to line EXP means polar optical phonon scattering (POP) is dominant. Combining the analysis of dislocation scattering mechanism in which the dislocation scattering is accompanied by polar optical phonon scattering, that is this charged impurity scattering above should be the manifestation of dislocation scattering. Therefore, the dislocation scattering should be the dominant CSM at the temperature range of 309K to 425K, and both the dislocation scattering and the polar optical phonon scattering are induced by the phase transition of α SnSe to β SnSe. The reason is that the sample (SnSe-HPS) has nanometer structure (less than 60 nm) [16]. It is this nanometer structure induces the phase transition of α SnSe to β SnSe occuring at lower temperature than 807K. The phase transition has been finished above 525K and acoustic phonon scattering become the dominant CSM. Therefore, polycrystal SnSe might have different CSM depending on the crystal structure. Besides, dislocation scattering and polar optical phonon scattering might be one of the approaches to improve thermoelectric property because the scattering factor for dislocation scattering (3/2) or polar optical phonon scattering (1/2) is larger than that of acoustic phonon scattering or AS [47].

## 5. Conclusion

Compared with the mobility-temperature dependent method, the electrical conductivity ratio (ECR) method used to analyze carrier scattering mechanism has a very unique advantage. Here, based on the relaxation time approximation, the applicability of ECR to complicated band structure is discussed. According to thermoelectric properties, the feasibility using ECR method of ideal single band transport model to study the carrier scattering mechanism of semiconductor materials with complicated band structure is studied. The difference between the carrier scattering mechanism for SnSe single crystal and polycrystals is analyzed systematically by ECR method using idea single-valley parabolic band model. The results can be described as below.

1. ECR method is not only applicable to idea band structure as reported previous but also to the complicated band structure, such as anisotropic band structure with single or multi-valley parabolic band and isotropic band structure with single or multi-valley nonparabolic band.
2. The analysis results of the carrier scattering mechanism of single crystal SnSe by ECR method using ideal single-valley parabolic model agree with the mobility-temperature dependence in the non-phase transition temperature range. Therefore, it is feasible to study the carrier scattering mechanism of semiconductor materials with complicated band structure by ECR method using ideal single-valley parabolic band transport model.
3. Dislocation scattering mechanism has very similar manifestation to charged impurity scattering. The difference between dislocation scattering and charged impurity scattering is that dislocation scattering is always accompanied by polar optical phonon scattering, such as the carrier scattering mechanism of SnSe along b axis direction at 573K to 823K. Therefore, compared with the mobility-temperature dependent method, the difference between ionized impurity scattering and dislocation scattering can be more easily distinguished by ECR method. The reason is that in the process of ECR analysis, it is always necessary to consider the role of multiple scattering mechanisms.
4. ECR method is a convenient method to identify the temperature dependence of carrier scattering mechanism (CSM), such as acoustical phonon scattering (APS), alloy scattering (AS), polar optical phonon scattering (POP), charged impurity scattering (CIS) and dislocation scattering (DS). The reason is that the temperature dependences of the electrical conductivity and Seebeck coefficient with good test reproducibility are easier to test than the Hall coefficient.
5. The temperature dependences of carrier scattering mechanisms along three direction of single crystal SnSe are different because of its anisotropic crystal structure. For polycrystalline SnSe, the carrier scattering mechanisms varies with the crystal structure and temperature too. For instance, in the case of vacuum-synthesized polycrystalline SnSe, both before and after the phase transition, acoustic phonon scattering is the dominant scattering mechanism. However, during the phase transition process (473K < T < 673K), polarization scattering is the dominant scattering mechanism. This result also indicates that the end temperature of the phase transition for this polycrystalline SnSe is lower than that of single-crystalline SnSe (823K). Besides, the nano-crystalline SnSe prepared by high-pressure sintering mainly exhibits ionized impurity scattering in the temperature range of 309K ≤ T ≤ 425K. When 425K < T ≤ 525K, dislocation scattering becomes the dominant mechanism. Moreover, its phase transition is completed at 525K.
6. Dislocation scattering and polar optical phonon scattering might be one of the approaches to improve thermoelectric property because the scattering factor for dislocation scattering (3/2) or polar optical phonon scattering (1/2) is larger than that of acoustic phonon scattering or alloy scattering.

**Acknowledgments**: This work was financially supported by The National Key Research and Development Program of China [Grand No. 2017YFF0204706], Overseas Expertise Introduction Project for Discipline Innovation or 111 Project (Grant No. X2017004), and the Fundamental Research Funds for the Central Universities [Grant No. FRF-MP-18-005 and FRF-MP-19-005]. We acknowledge discussions with Dr. Jian He.

**Reference**
1. Guiying Xu, Pan Ren, Tie Lin, Xiaofeng Wu, Yanhua Zhang, Sitong Niu, Trevor P. Bailey, Mechanism and Application Method to Analyze Carrier Scattering Factor by Electrical Conductivity Ratio Based on Thermoelectric Property Measurement, Journal of Applied Physics,J. Appl. Phys. 123, 015101 (2018)
2. Tao Mao, Pan Ren, Guiying Xu, Junling Gao, Jiaxin Di, Libo Zhang, Thermoelectric performance of p-type Nd1-xInxTe3 fabricated by high pressure sintering method, Journal of Alloys and Compounds, 696 (2017), 648-654.
3. S. M. A. Taher, J. B. Gruber, and L. C. Olsen. Electrical transport properties of single crystal rare earth sesquisulfides, J. Chem. Phys.60 (1974), 5, 2050-2056.
4. Yubo Luo, Junyou Yang, Qinghui Jiang, Weixin Li, Dan Zhang, Zhiwei Zhou, Yudong Cheng, Yangyang Ren, Xu He,. Investigation on the microstructure and thermoelectric performance of magnetic ions doped Bi0.5Sb1.5Te3 solidified under a magnetostatic field. Acta Materialia, 127, 185-191, 2017
5. Wen Li, Siqi Lin, Binghui Ge, Jiong Yang, Wenqing Zhang, and Yanzhong Pei. Low Sound Velocity Contributing to the High Thermoelectric Performance of Ag8SnSe6. Adv. Sci., 3, 1600196, 2016.
6. Mahn Jeong, Jang-Yeul Tak, Soonil Lee, Won-Seon Seo, Hyung Koun Cho, Young Soo Lim. Effects of Cu incorporation as an acceptor on the thermoelectric transport properties of CuxBi2Te2.7Se0.3 compounds. Journal of Alloys and Compounds, 696, 213, 2017.
7. Changho Lee, Yong-Ho Park, and Hitoshi Hashimoto. Effect of nonstoichiometry on the thermoelectric properties of a Ag2Se alloy prepared by a mechanical alloying process. Journal of Applied Physics, 101, 2, 024920, 2007.
8. M. Gao, J. Zhang and D. M. Rowe, Thermoelectric Transformation and Application, Weapon Industry Press, Beijing, 1995, pp. 34-41.
9. Y. Nishio and T. Hirano, Improvement of the efficiecy of the thermoelectric energy conversion by utilizing potential barriers, Jpn. J. Appl. Phys., Part 36, 170 (1997).
10. A. Popescu, L. M. Woods, J. Martin, and G. S. Nolas, Model of transport properties of thermoelectric nanocomposite materials, Physical Review B 79, 205302 _2009_
11. Y. Kajikawa, Effects of potential barrier height and its fluctuations at grain boundaries on thermoelectric properties of polycrystalline semiconductors, Journal of Applied Physics 114, 053707 (2013).
12. Xinhong Guan, Pengfei Lu, Liyuan Wu, Lihong Han, Gang Liu, Yuxin Song, Shumin Wang, Thermoelectric properties of SnSe compound, Journal of Alloys and Compounds, 643 (2015) 116–120.
13. Li-Dong Zhao, Shih-Han Lo, Yongsheng Zhang, Hui Sun, Gangjian Tan, Ctirad Uher, C. Wolverton, Vinayak P. Dravid, & Mercouri G. Kanatzidis1, Ultralow thermal conductivity and high thermoelectric figure of merit in SnSe crystals, Nature, Vol 508, 17 April2014, 373.
14. M. M. Nassary, The electrical conduction mechanisms and thermoelectric power of SnSe single crystals, Turk J Phys, 33 (2009) , 201 – 208.
15. J.C. Li, D. Li, X.Y. Qing, J. Zhang, Enhanced thermoelectric performance of p-type SnSe doped with Zn, Scripta Materialia 126 (2017) 6–10.
16. Junling Gao, Huina Zhu, Tao Mao, Libo Zhang, Jiaxin Di, Guiying Xu, The effect of Sm doping on the transport and thermoelectric of SnSe, Materials Research Bulletin 93 (2017) 366–372.

17. Junling Gao, Guiying Xu, Thermoelectric performance of polycrystalline Sn1-xCuxSe (x = 0–0.03), prepared by high pressure method, Intermetallics 89 (2017) 40–45.
18. Yue-Xing Chen, Zhen-Hua Ge, Meijie Yin, Dan Feng, Xue-Qin Huang, Wenyu Zhao, and Jiaqing He, Understanding of the Extremely Low Thermal Conductivity in High-Performance Polycrystalline SnSe through Potassium Doping, Adv. Funct. Mater. 2016, 26, 6836–6845.
19. Junling Gao, Ziming Shao, Guiying Xu, Thermoelectric transport properties of Sn1-xGexSe (x=0-0.03) prepared by melting synthesis method, Int J Appl Ceram Technol. 2017;14:963–968.
20. WANG Y, BAI S, SHI H, et al. Modulating structures to decouple thermoelectric transport leads to high performance in Polycrystalline Sn Se[J]. Journal of Materials Chemistry A, 2024, 12: 144-152.
21. LI S, HOU Y, ZHANG S, et al. Realizing high thermoelectric performance in magnetic field-assisted solution synthesized nanoporous SnSe integrated with quantum dots[J]. Chemical Engineering Journal, 2023, 451: 138637.
22. HUANG X, GONG Y, LIU Y, et al. Achieving high isotropic figure of merit in Cd and in codoped polycrystalline Sn Se[J]. ACS Applied Materials & Interfaces, 2024, 16: 20400-20408.
23. MANDAL P, MAITRA S, GHORUI U K, et al. Effects of codoping on tin selenide nanomaterials to enhance the thermoe-lectric performance above the ambient temperature range[J]. Journal of Materials Chemistry C, 2023, 11: 8577-8589.
24. QIN Y, XIONG T, ZHU J, et al. Realizing high thermoelectric performance of Cu and Ce co-doped p-type polycrystalline SnSe via inducing nanoprecipitation arrays[J]. Journal of Advanced Ceramics, 2022, 11: 1671-1686.
25. GONG Y, ZHANG S, HOU Y, et al. Enhanced density of states facilitates high thermoelectric performance in solution-Grown Ge- and In-Codoped Sn Se nanoplates[J]. ACS Nano, 2022, 17: 801-810.
26. Qin B, Wang D, Hong T, et al. High thermoelectric efficiency realized in Sn Se crystals via structural modulation. Nat Commun, 2023, 14: 1366.
27. Xu P, Hua Y, Jin K, et al. Heavy Rare Earth Element Gd Enhancing Thermoelectric Performance in p-Type Polycrystalline Sn Se via Optimizing Carrier Transport and Density of States [J]. Chemical Research in Chinese Universities, 2023, 39(4): 690-696.
28. Zhu J, Bo L, Kong J, et al. Enhanced thermoelectric and mechanical properties of polycrystalline cubic Sn Se by Ag Bi Te2 alloying [J]. Journal of Alloys and Compounds, 2024, 971: 172754.
29. Lu W, Wu S, Ding Q, et al. Cold Sintering Mediated Engineering of Polycrystalline Sn Se with High Thermoelectric Efficiency [J]. ACS Applied Materials & Interfaces, 2024, 16(4): 4671-4678.
30. Long Zhang, Lei Lu, Dechao Zhang, Wentao Hu, Ning Wang, Bo Xu, Yueming Li, Hong Zeng, Dual-buffered SnSe@CNFs as negative electrode with outstanding lithium storage performance Electrochimica Acta 209 (2016) 423–429.
31. Zhian Zhang, Xingxing Zhao, Jie Li, SnSe/carbon nanocomposite synthesized by high energy ball milling as an anode material for sodium-ion and lithium-ion batteries, Electrochimica Acta 176 (2015) 1296–1301.
32. Babu Pejjai, Vasudeva Reddy, Minnam Reddy, Sreedevi Gedi, Chinho Park, Status review on earth-abundant and environmentally green Sn-X (X ¼ Se, S) nanoparticle synthesis by solution methods for photovoltaic applications, International Journal of Hydrogen Energy 42 ( 2017 ) 2790-2831.
33. Min Jin, Zhiwei Chen, Xiaojian Tan, Hezhu Shao, Guoqiang Liu, Haoyang Hu, Jingtao Xu, Bo Yu, Hui Shen, Jia-Yue Xu, Haochuan Jiang, Yanzhong Pei, and Jun Jiang. Charge Transport in Thermoelectric SnSe Single Crystals. ACS Energy Letters, 3, 3, 689-694. 2018.
34. Volker Schmidt, Philipp F. J. Mensch, Siegfried F. Karg, Bernd Gotsmann, Pratyush Das Kanungo, Heinz, Schmid, and Heike Riel, Using the Seebeck coefficient to determine charge carrier concentration, mobility, and relaxation time in InAs nanowires, Appl. Phys. Lett. 104, 012113 (2014).
35. Y. Kajikawa, K. Okamura, Y. Inoko, and H. Mizuki, Analysis of temperature dependence of electrical

conductivity in degenerate n-type polycrystalline InAsP films in an energy-filtering model with potential fluctuations at grain boundaries, Journal of Applied Physics 112, 123712 (2012).
36. Guangsha Shi and Emmanouil Kioupakis, Quasiparticle band structures and thermoelectric transport properties of p-type SnSe, Journal of Applied Physics 117, 065103 (2015).
37. Karlheinz Seeger, Semiconductor Physics, , Springer Study Edition., Springer-Verlag Berlin Heidelberg: Vienna and New York, 1973.
38. B. Pödör, Eleotron mobility in plastioally deformed germanium, Phys. Stat. Sol. 16 (1966), K167.
39. J. H. P. van Weeren, R. Struikmans, and J. Blok, Test of the read model concerning the electron mobility in plastically deformed n-type Ge, Phys. Stat. Sol. 19 (1967), K107.
40. A. F. Gibson, The mobility, diffusion constant, and lifetime of minority carriers in heavily dislocated germanium, J. Phys. Chem. Solids, 8 (1959), 147.
41. Cheng-Lung Chen, Heng Wang, Yang-Yuan Chen, Tristan Day, and G. Jeffrey Snyder，Thermoelectric properties of p-type polycrystalline SnSe doped with Ag, Journal of Materials Chemitry A, 2014, 2, 11171
42. J.Q. Li, L.F. Li, S.H. Song, F.S. Liu, and W.Q. Ao,. High thermoelectric performance of GeTe–Ag8GeTe6 eutectic composites. J. Alloys Compd. 144 (2013), 565.
43. A. Okazaki, I. Ueda. The crystal structure of stannous selenide SnSe. J. Phys. Soc. Jpn. 11 (1956), 470.
44. H. Wiedemeier, H.G. Schnering, Zeitschrift fur Kristallographie. Refinement of the structures of GeS, GeSe, SnS and SnSe. 148 (1978), 295-303.
45. Y. L. Li, X. Shi, D.D. Ren, J.K. Chen, L.D. Chen,. Investigation of the anisotropic thermoelectric properties of oriented polycrystalline SnSe. Energies, 8 (2015), 6275-6285.
46. B. Predel, O. Madelung,,  H. Landolt, Richard B Ornstein, & K. H. Hellwege, Phase Equilibria, Crystallographic amd Thermodynamic Data of Binary Alloys. Springer-Verlag, 1991..
47. A. F. Ioffe, Semiconductor thermoelement and thermoelectric cooling, London, Infosearch, 1957.